\documentclass{article}
\usepackage{ital-ia}
\usepackage[utf8]{inputenc}
\usepackage[T1]{fontenc}
\usepackage{color}
\usepackage{url}

\usepackage{times}
\usepackage{soul}
\usepackage{url}
\usepackage[hidelinks]{hyperref}
\usepackage[utf8]{inputenc}
\usepackage[small]{caption}
\usepackage{graphicx}
\usepackage{amsmath}
\usepackage{booktabs}
\usepackage{algorithm}
\usepackage{algorithmic}
\hypersetup{colorlinks=true,linkcolor=blue,citecolor=blue,urlcolor=blue}
\newcommand{\orcid}[1]{\href{https://orcid.org/#1}{\includegraphics[width=8pt]{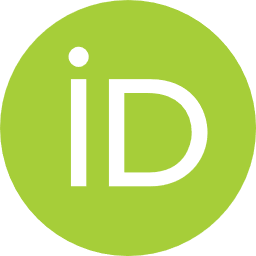}}}

\title{Responsible AI in Healthcare\footnote{Presented at Ital-IA 2022: The 2nd CINI National Conference on Artificial Intelligence, February 9--11, 2022, Turin, Italy (\emph{online event}). URL: \url{https://www.ital-ia2022.it/}}\\\Large Overview of the Research Activities carried out at the Department of Informatics, Systems, and Communication of the University of Milano-Bicocca, within the CINI National Laboratory\\``Artificial Intelligence and Intelligent Systems'' (AIIS)\footnote{The National Laboratory \emph{Artificial Intelligence and Intelligent Systems} (AIIS) was created by the \emph{Consorzio Interuniversitario Nazionale per l'Informatica -- National Interuniversity Consortium for Informatics} (CINI) in June 2018 with the support of the \emph{Security Intelligence Department} of the \emph{Presidency of the Council of Ministers} to develop common objectives between public institutions, Italian industry and scientific research of national universities and research centers. URL: \url{https://www.consorzio-cini.it/index.php/en/labaiis-home}}}

\author{
    Federico Cabitza$^\diamond$\orcid{0000-0002-4065-3415}, Davide Ciucci$^\diamond$\orcid{0000-0002-8083-7809}, Gabriella Pasi$^\star$\orcid{0000-0002-6080-8170}, and Marco Viviani$^\star$\orcid{0000-0002-2274-9050}
    \affiliations
    \large
    Università degli Studi di Milano-Bicocca\\Dipartimento di Informatica, Sistemistica e Comunicazione (DISCo)\\Edificio U14 (ABACUS), Viale Sarca, 336 -- 20126 Milan, Italy\\
    $^\diamond$MUDILAB: Modeling Uncertainty, Decisions and Interaction\\
    Website: \url{https://mudilab.disco.unimib.it/}\\
    $^\star$IKR3 LAB: Information and Knowledge Representation, Retrieval, and Reasoning\\
    Website: \url{https://ikr3.disco.unimib.it/}
    \emails
    \texttt{\small \{federico.cabitza, davide.ciucci, gabriella.pasi, marco.viviani\}@unimib.it}
}

\begin{document}

\maketitle

\begin{abstract}
This article discusses open problems, implemented solutions, and future research in the area of responsible AI in healthcare. In particular, we illustrate two main research themes related to the work of two laboratories within the Department of Informatics, Systems, and Communication at the University of Milano-Bicocca. The problems addressed concern, in particular, {uncertainty in medical data and machine advice}, and the problem of online health information disorder.
\end{abstract}

\section{Research Themes}

According to the \emph{Ethics Guidelines for Trustworthy Artificial Intelligence} (AI) \cite{euTrust2021}, a document defined by the \emph{High-Level Expert Group on Artificial Intelligence} (AI HLEG) set up by the European Commission, seven are the key requirements that AI systems should meet in order 
to be trustworthy:
\begin{enumerate}
    \item \emph{Human agency and oversight}, i.e., supporting human autonomy and decision-making, as prescribed by the principle of respect for human autonomy;
    \item \emph{Technical robustness and safety}, i.e., including resilience to attack and security, fall back plan and general safety, accuracy, reliability and reproducibility;
    \item \emph{Privacy and data governance}, i.e., including respect for privacy, quality and integrity of data, and access to data;
    \item \emph{Transparency}, i.e., including traceability, explainability and communication;
    \item \emph{Diversity, non-discrimination and fairness}, i.e., including the avoidance of unfair bias, accessibility and universal design, and stakeholder participation;
    \item \emph{Societal and environmental well-being}, i.e., including sustainability and environmental friendliness, social impact, society and democracy;
    \item \emph{Accountability}, i.e., including auditability, minimisation and reporting of negative impact, trade-offs and redress.
\end{enumerate}

The researchers of the Department of Informatics, Systems and Communication (DISCo) of the University of Milano-Bicocca, who work on the macro theme of the article, are organized in different research laboratories active on two main research themes, addressing some of the key requirements illustrated before in the health domain.
In particular, Federico Cabitza and Davide Ciucci, members of the \emph{Modeling Uncertainty, Decisions and Interaction Laboratory} (MUDILAB), discuss the problem of uncertainty in data that feed machine learning algorithms and the importance of the cooperation between AI and human decision-makers in healthcare. All the seven key requirements are involved, with particular reference to 1, 4, and 6.
Gabriella Pasi and Marco Viviani, members of the \emph{Information and Knowledge Representation, Retrieval and Reasoning Laboratory} (IKR3 LAB), address the problem of health information disorder and discuss several open issues and research directions related mainly to key requirements 1, 2, 4, 5, and 6.

\section{Responsible AI as a Support for Healthcare Decisions (F. Cabitza and D. Ciucci)}

In this section, we address some responsibility and trustworthiness issues in current machine learning algorithms
and decision support systems in healthcare. First of all, medical data can be affected by different types of uncertainty/variability, some of which are not usually accounted for when developing ML models. In particular, we refer to different forms of variability:
\begin{itemize}
\item \emph{Biological variability}, which occurs when a person is associated with more or less slightly different values that express a health condition over time;
\item \emph{Analytical variability}, which occurs when a testing equipment, although calibrated, produces different values for a specific patient/subject with respect to other equipment (from the same vendor or different vendors);
\item \emph{Pre-} and \emph{post-analytical variability}, which occurs when different values in the same exam for the same subject can be due to different ways (including erroneous ones) to use the equipment or produce data about test results.
\end{itemize}
These sources of variability add on to the noise due to more common (and treated) sources of data or label noise \cite{cabitza2016information,CCR19,CLARCB19,HW21}:
\begin{itemize}
\item \emph{Missing data}, in different forms, e.g., a value that is not known or a patient that does not reveal a symptom;
\item \emph{Vagueness}, such as a symptom is {\em mild} rather than {\em severe};
\item A physician undecided on the \emph{interpretation} of an exam, perhaps with a degree of confidence;
\item \emph{Noise}, in instruments or in reporting data.
\end{itemize}
 In light of these considerations, it is important to get awareness of potential sources of noise in biological and clinical data and conceive novel methods to both mitigate their impact and manage the related variability and uncertainty.

To this aim, we are developing a set of new algorithms able to cope with all these flaws in healthcare data. We explore different approaches: \emph{partially labeled data} \cite{CCC20}, \emph{superset learning} \cite{CCH21}, \emph{multi-rater annotation} \cite{CCSFC21}, \emph{cautious learning} \cite{CCBC21}, and \emph{soft clustering} \cite{CC19}. 
The goal is to  create a framework for robustness validation of classification systems based on Machine Learning. The developed tools and algorithms should be able to handle different forms of uncertainty simultaneously and to abstain from giving a precise answer whenever this is not possible or too risky.
 
 Moreover, we advocate the need to move beyond aggregation methods by mere majority voting in ground truthing~\cite{CCBC21}, that is the production of the ground truth labels to be used in supervised learning, as this could result in excessive simplification with respect to the complexity of the phenomenon at hand, for which multiple right and complementary interpretations are possible to coexist for a single case~\cite{BCC21}.

Finally, we also advocate further research on the design and evaluation of alternative interaction protocols~\cite{CCS21} stipulating how human decision makers could use, and in some case even collaborate, with AI-based decision support systems, in order to mitigate the risk of having cognitive biases, like automation bias, automation complacency, AI over-reliance and its opposite, the prejudice against the machine~\cite{Cab19}, which undermine the effectiveness and efficiency of computer-supported decision making process.
This will lead to more reliable and trustworthy decision support systems.

\section{Responsible AI and Health Information Disorder (G. Pasi and M. Viviani)}

In this section we address the issue of the responsibility and trustworthiness of AI algorithms in the context of the spread of different forms of health-related communication pollution. This is an important issue, which is fundamental to both understand and limit the  generation and diffusion of \emph{rumors}, \emph{misinformation}, and \emph{disinformation} \cite{guess2020misinformation,wardle2018thinking}, especially in the health domain \cite{DiSotto2022,swire2019public,viviani2017credibility}. All these forms of false, unreliable, low-quality information, generated with or without fraudulent intent, have recently been grouped under the name of \emph{information disorder} \cite{wardle2017information}.

\subsection{Health Information Disorder Generation}

A first aspect that needs to be addressed in this context is 
that there are several systems based on AI techniques that have allowed in recent years: $(i)$ the generation of increasingly realistic fraudulent content, and $(ii)$ the large-scale dissemination of the same, often, with manipulative intent \cite{bontridder2021role}. As far as the first aspect is concerned, let us think, for example, to the phenomenon of \emph{deep fakes} \cite{hancock2021social};  as far as the second aspect is concerned, we may cite the increasing effectiveness of the systems of \emph{micro-targeting} \cite{zuiderveen2018online}, of \emph{information filtering} \cite{chitra2020analyzing}, and of \emph{social bot generation} \cite{allem2018could}.

The ethical implications of information disorder generation 
essentially concern \emph{human dignity}, \emph{autonomy}, and \emph{democracy} \cite{baeroe2020achieve,bontridder2021role}. Human dignity, because people are treated not as persons but as ``temporary aggregates of data processed at industrial scale'' \cite{euTrust2021}, often with opinion manipulation intents, leaving to people the impression that they are receiving the same information as any other person in the digital ecosystem when in fact they are part of \emph{filter bubbles} \cite{holone2016filter}. This problem is closely related to that of autonomy, in fact users are not completely able to build their own (digital) identity. Finally, manipulation leading to excessive polarization (as seen above) produces impossibility to make globally shared decisions, leading to serious repercussions in several areas of well-being, not least health.

In this area, we are working on the definition of models and methodologies based on graph mining and NLP techniques for the identification of \emph{echo chambers} and limiting the problem of their formation, which is closely related to the problem of selective filtering of information, including with respect to the health domain \cite{villa2021echo}.

\subsection{Health Information Disorder Detection}

Due to the spread of online information disorder, a second aspect that needs to be addressed is to identify different forms of communication pollution in different media formats, including in the health domain; in this context several solutions have been proposed in recent years \cite{cui2020deterrent,dharawat2020drink,hou2019towards,upadhyay2021health,zhao2021detecting}.

\textcolor{black}{Ethical issues that arise in this area concern the fact that the algorithms developed to identify information disorder should not impede \emph{freedom of expression} and \emph{autonomy} in decision making. According to \cite{brachman-schmolze:kl-one}, ``permitting AI systems to regulate content automatically
would [...] seriously affect freedom of expression and information''. This, in particular, because ``it is not clear how often and under which circumstances \emph{ex ante} filtering or blocking take place'' \cite{marsden2019regulating}, and because AI systems trained to detect information disorder could produce false positives and false negatives. Indeed, such systems
``could lead to over-censorship of legitimate content that is machine-labeled incorrectly as
disinformation'' \cite{marsden2019regulating}. Such problems, related to a non-transparent or incorrect identification or filtering of information judged as not genuine, brings with it the problems related to awareness, and therefore autonomy, in decisions.}

\textcolor{black}{Therefore, in the development of solutions for the identification of information disorder (also related to health information), we are currently investigating models and methodologies that allow users not to have a \emph{hard filter} with respect to access to information (based on their genuineness estimated by the system). In fact, we are evaluating the possibility of providing users with a \emph{ranking} of the information that takes into account a \emph{gradual} notion of genuineness \cite{goeuriot2021clef,putri2021multi}, instead of a binary notion as done so far in the literature.}

\textcolor{black}{Other issues concern: $(i)$ \emph{data collection} and \emph{data processing}, since they can be carried out incorrectly, or on data that already contain bias, or are incomplete with respect to different cultural environments (e.g., based on the use of a single language)}; $(ii)$ the presence of \emph{opacity} in algorithms with respect to not obvious connections between the data used, how they were used, and the obtained conclusions, which, in addition, can only be as reliable as the data they are based on. Such issues lead to the so-called \emph{inscrutable evidence} and \emph{misguided evidence}, as reported in \cite{trocin2021responsible}.

\textcolor{black}{To take these issues into consideration, we have recently worked on the definition suitable labeled datasets and evaluation strategies within the CLEF \emph{eHealth Evaluation Lab} \cite{goeuriot2021clef}, which focuses on the \emph {Consumer Health Search} (CHS) task \cite{goeuriot2021consumer}}.
In addition, we have been working on the development of model-driven solutions for the evaluation of the genuineness of information (including health information), which are based on the use of \emph{Multi-Criteria Decision Making} (MCDM) techniques where the system proves to be explainable with respect to the results obtained \cite{pasi2019multi}.

\section{Funded Research Projects}

\begin{itemize}
    \item DoSSIER. Horizon 2020 ITN on \emph{Domain Specific Systems for Information Extraction and Retrieval} 
    (H2020-EU.1.3.1., ID: 860721). URL: \url{https://dossier-project.eu/}
\end{itemize}

\section{Organized Research Initiatives}

\begin{itemize}
    \item CLEFeHealth 2021 -- Task 2: Consumer Health Search. Evaluation Lab held in conjunction with CLEF 2021: The 12th Conference and Labs of the Evaluation Forum. September 21--21, 2021, Bucharest, Romania (\emph{online event}). URL: \url{https://clefehealth.imag.fr/?page_id=610}
    \item IPMU 2022: The 19th Conference on Information Processing and Management of Uncertainty in Knowledge-Based Systems. July 11--15, 2022, Milan, Italy.\\URL: \url{https://ipmu2022.disco.unimib.it/}
    \item ROMCIR 2022: The 2nd Workshop on Reducing Online Misinformation through Credible Information Retrieval. Held in conjunction with ECIR 2022: The 44th European Conference on Information Retrieval. April 10--14, 2022,  Stavanger, Norway.\\URL: \url{https://romcir2022.disco.unimib.it/}
    \item TrueHealth 2022:
Special Track on Combating Health Information Disorder in the Digital World for Social Wellbeing. Held in conjunction with GoodIT 2022: The ACM International Conference on Information Technology for Social Good. September 7--9, 2022 Limassol, Cyprus. URL: \url{https://truehealth.disco.unimib.it/}
\end{itemize}

\section{Applications}

\begin{itemize}
    \item A set of ML tools to manage uncertainty:
    \begin{itemize}
        \item \url{https://github.com/AndreaCampagner/scikit-weak/} 
        \item \url{https://github.com/AndreaCampagner/uncertainpy/}
    \end{itemize}
    \item An approach for assessing the presence of online echo chambers (including in healthcare):
    \begin{itemize}
        \item \url{https://github.com/ikr3-lab/echochambers/}
    \end{itemize}
    \item A data science solution for assessing the genuineness of health information disseminated in the form of Web pages and social media content:
    \begin{itemize}
        \item \url{https://github.com/ikr3-lab/health-misinformation/}
    \end{itemize}
\end{itemize}

\section{Challenges and Perspectives}
\begin{itemize}
    \item In a continuously situation of growing attention and  faith in AI tools for healthcare decision making, our approach is aimed to frame the usefulness and limits of these tools. The proposed decision support system, with related components, will improve the trustworthiness and robustness of the  decisions taken by  physicians and generally by healthcare decision-makers.
    \item \textcolor{black}{Concerning responsible AI with respect to health information disorder, a fundamental role will be played by the study and development of solutions that are as transparent as possible and understandable to users, together with preserving confidentiality \cite{livraga2019data}, especially in a domain as sensitive as health. In this area, in particular, it will be necessary to deepen how to take into account a correct domain knowledge also with the help of medical experts. From the technological point of view, this will be achieved through the development of hybrid models that are partly model-driven and partly data-driven.}
\end{itemize}

\bibliographystyle{named}
\bibliography{ital-ia}

\end{document}